\documentclass[page-classic]{epl2} 
\usepackage{amssymb}

\title{Oscillation of the spin-currents of cold atoms on a ring due
to light-induced spin-orbit coupling}
\shorttitle{Oscillation of the spin-currents of cold atoms on a ring} 

\author{Wenfang Xie\inst{1} \and Yanzhang He\inst{2} \and Chengguang Bao\inst{2}\thanks{The corresponding author: stsbcg@mail.sysu.edu.cn}}

\institute{
  \inst{1} School of Physics and Electronic Engineering, Guangzhou University, Guangzhou 510006, People's Republic of China\\
  \inst{2} School of Physics and Engineering, Sun Yat-Sen University, Guangzhou 510275, People's Republic of China
}
\pacs{03.75.Kk}{Dynamic properties of condensates; collective and hydrodynamic excitations, superfluid flow}
\pacs{03.75.Mn}{Multicomponent condensates; spinor condensates}
\pacs{03.75.Nt}{Other Bose-Einstein condensation phenomena}

\abstract{
The evolution of two-component cold atoms on a ring with quasispin-orbit
coupling (qSO) and spin-flip has been studied analytically for the case with
N noninteracting particles. Then, the effect of interaction is evaluated
numerically via a two-body system. Oscillating persistent spin-currents have
been found, and the underlying regularity governing the period and amplitude of the
oscillation has been unveiled. A set of formulae have been derived to describe the
oscillation so that the underlying physics can be understood in an analytical way.
Based on these formulae the oscillation can be better controlled via adjusting the
parameters of the laser beams.
}

\begin{document}

\maketitle

It is well known that the study of the motion of charged particles under a
magnetic field is an essential topic in both macroscopic and microscopic
physics. In particular, a number of distinguished quantum mechanic
phenomena, such as the Aharonov-Bohm (A-B) oscillation and the fractional
quantum Hall effect (FQHE), are caused by the magnetic gauge
field.\cite{hz,ya,dct,rbl} After the experimental realization of the
condensation of neutral atoms with nonzero spin,\cite{jst} a great interest
is to create a light-induced gauge vector field so that various
magnetic-electronic phenomena in condense matter can be copied in the
condensates of neutral atoms.\cite{hz,slz} In recent years, by making use of
the polarized laser beams, effective Lorentz force imposed on the
multi-component neutral cold atoms can be created which leads to
quasispin-orbit coupling (qSO).\cite{yjl1,yjl2,yjl3,yjl4,sb,jyz} This
technique opens a new perspective in the field of BEC. In particular, the
qSO will lead to the spin-Hall effect characterized by the persistent
spin-currents.\cite{slz,yjl1} Recently, the oscillation of the spin-currents
has been observed in a few experiments.\cite{yjl1,slz,jyz} However, the
underlying regularity has not yet been well understood.

On the other hand it is now possible to trap a condensate in a ring
geometry. Long-lived rotational superflows have been induced in this kind of
systems.\cite{sb,sg,asa,cr,kh,ar,bes,sm,kcw} Making use of the laser beams
the creation and observation of the spin-currents on a ring has been
experimentally realized.\cite{sb} It was found that the stability depends
strongly on the initial ratio of the two components.\cite{sb}, The
underlying regularity remains to be studied.

This paper is dedicated to the two-component condensates on a ring under
qSO. The aim is to clarify the regularity governing the oscillation of the
spin-currents. The ring is considered as one-dimensional. When the
interaction is not taken into account, analytical solution can be obtained
so that the oscillation can be understood in an analytical way. The emphasis
is placed on unveiling the regularity governing the period and amplitude of
oscillation. Finally, the effect of the interaction is evaluated via a
two-body system.

Let the quasi-spin $\hat{s}$ be introduced to describe the two
components of an atom as usual. The state with $s_z=1/2\
(-1/2)$ is named the up- (down-) state denoted as
$\psi_{\uparrow}$ and $\psi_{downarrow}$, respectively. These
two states can be transformed to each other via the Raman
coupling. When two counter-propagating and polarized laser
beams are applied, the two components of atoms will move
towards opposite directions along the beams. The interaction is
firstly neglected, its effect is evaluated later. Then, the
hamiltonian is just $H=\Sigma_i\hat{h}_i$, where $\hat{h}_i$ is
for the $i$-th particle. We define a $U$-transformation so that
$\phi_{\uparrow}=e^{i\beta\theta}\psi_{\uparrow}$ and
$\phi_{\downarrow}=e^{-i\beta\theta}\psi_{\downarrow}$. For a
one-dimensional ring the Hamiltonian for $\phi_{\uparrow}$ and
$\phi_{\downarrow}$ is \cite{hz,wz,yl,slz,yjl1}
\begin{equation}
 U\hat{h}U^{-1}
  =  (-i
       \sigma_I
       \frac{\partial}{\partial\theta}
      -\beta
       \sigma_z)^2
    +\delta
     \sigma_z
    +\gamma
     \sigma_x
\end{equation}
where $\theta$ is the azimuthal angle along the ring. The unit of energy is
hereafter $E_{unit}\equiv\hbar^2/(2mR^2)$, where $m$ is the mass of
an atom, $R$ is the radius of the ring. $\sigma_I$ is just a unit matrix
with rank 2. $\beta\equiv k_0 R$, where $2k_0$ is the momentum transfer
caused by the two lasers. $\gamma\equiv\frac{\Omega}{2E_{unit}}$ , where
$\Omega/2$ is the strength of Raman coupling causing the spin-flips.
$\delta\equiv\frac{\delta'}{2E_{unit}}$ is the Raman detuning, where
$\delta'=\varepsilon_{split}-\hbar\omega _{\delta}$, $\varepsilon_{split}$
is the Zeeman energy difference between the two
spin-states, and $\omega_{\delta}$ is the frequency difference between
the two laser beams. Note that, due to the ring geometry, $\beta$ must be
an integer. It implies that the transfer of momentum will be suppressed
unless the momentum transfer is\ close to a specific set of values.

$\hat{h}$ has two groups of eigenstates, they can be written as
\cite{wz,yl,yjl1}
\begin{eqnarray}
 \psi_k^{(+)}
 &=& \sin(\rho_k)
     e^{-i\beta\theta}
     \varphi_{k\uparrow}
    +\cos(\rho_k)
     e^{i\beta\theta}
     \varphi_{k\downarrow},  \nonumber \\
 \psi_k^{(-)}
 &=& \cos(\rho_k)
     e^{-i\beta\theta}
     \varphi_{k\uparrow}
    -\sin(\rho_k)
     e^{i\beta\theta}
     \varphi_{k\downarrow },
\end{eqnarray}
where
\begin{equation}
 \varphi_{k\uparrow}
  =  \frac{1}{\sqrt{2\pi}}
     e^{ik\theta}
     \left\{
      \begin{array}{c}
       1 \\
       0
      \end{array}
     \right\},\ \ \ \ \ \
 \varphi_{k\downarrow}
  =  \frac{1}{\sqrt{2\pi}}
     e^{ik\theta}
     \left\{
      \begin{array}{c}
       0 \\
       1
      \end{array}
     \right\}.
\end{equation}
$k$ is an $\pm$ integer, $\sin\rho_k=s_{\gamma}\sqrt{(a_k-2k\beta+\delta)/(2a_k)}$,
$\cos\rho_k=\sqrt{(a_k+2k\beta-\delta)/(2a_k)}$, $s_{\gamma}$ is the sign of
$\gamma$, $a_k=\sqrt{(2k\beta-\delta)^2+\gamma^2}$, and $\rho_k$ is ranged from
$-\pi/2\rightarrow\pi/2$. Note that $\rho_k$ can be rewritten as a function of
$|\gamma/(\delta-2k\beta)|$ together with the signs of $\gamma$ and $\delta-2k\beta$.
This feature is useful in the following discussion. The
eigenenergies of $\psi_k^{(\pm)}$ are $E_k^{(\pm)}=k^2+\beta^2\pm a_k$. Obviously,
$E_k^{(-)}\leq E_k^{(+)}$. Besides, $E_k^{(\pm)}=E_{-k}^{(\pm)}$ if $\delta=0$.

The time-dependent solution $\psi(\theta,t)$ of the single-particle
Schr\"{o}dinger equation starting from an initial state $\psi_{init}$ can be
formally written as
\begin{equation}
 \psi(\theta,t)
  =  e^{-i\tau\hat{h}}
     \psi_{init}
  =  \sum_{k\lambda}|
     \psi_k^{(\lambda)}\rangle
     e^{-i\tau E_k^{(\lambda)}}\langle
     \psi_k^{(\lambda )}|
     \psi_{init}\rangle,
\end{equation}
where $\tau\equiv tE_{unit}/\hbar$ (for $^{87}$Rb and $R=12\mu m$ as given
in \cite{sb}, $t=0.398\tau\sec$), $\lambda=\pm$.

In the experiment reported in \cite{jyz} the evolution was
induced by a sudden change of the laser field, namely, by
changing $\delta$ and/or $\gamma$. In this experiment a strong
magnetization oscillation is reported. To give a better
insight, it is assumed that a set of $\beta$, $\gamma_1$, and
$\delta_1$ are given, and the system is initially at the ground
state (g.s.), namely,
$\psi_{init}=\psi_{gs}=\cos(\rho_q^{(1)})e^{-i\beta\theta}
\varphi_{q\uparrow}-\sin(\rho_q^{(1)})e^{i\beta\theta
}\varphi_{q\downarrow}$. Where $q$ depends on the three
parameters so that the associated energy $E_q^{(-)}$ is the
lowest. The superscript in $\rho_q^{(1)}$ implies that this
quantity is calculated from the first set of parameters. Then,
$\gamma_1$ and $\delta_1$ are changed to $\gamma_2$ and
$\delta_2$ suddenly. Accordingly, we have a new set of
eigenstates. With them $\psi(\theta,t)$ becomes
\begin{equation}
 \psi(\theta,t)
  =  e^{-i\tau(q^2+\beta^2)}
     ( f_u\varphi_{q-\beta,\uparrow}
      -f_d\varphi_{q+\beta,\downarrow}),
\end{equation}
where $f_u=\cos(\rho_q^{(1)})\cos(a_q^{(2)}\tau)+i\cos(2\rho_q^{(2)}
-\rho_q^{(1)})\sin(a_q^{(2)}\tau)$, and $f_d=\sin(\rho_q^{(1)})
\cos(a_q^{(2)}\tau)+i\sin(2\rho_q^{(2)}-\rho_q^{(1)})\sin(a_q^{(2)}\tau)$.
Where $a_q^{(2)}$\ and $\rho_q^{(2)}$ are calculated from $\gamma_2$
and $\delta_2$. Let the time-dependent densities of the up- and
down-component be defined from the identity $\psi^{+}(\theta,t)\psi(\theta,t)
\equiv n_{\uparrow}(\theta,\tau)+n_{\downarrow}(\theta,\tau)$. Then, we have
\begin{eqnarray}
 n_{\uparrow}
 &=& \frac{1}{2\pi}
     \{\cos^2(\rho_q^{(1)})
    +[ \cos^2(2\rho_q^{(2)}
      -\rho_q^{(1)})
      -\cos^2(\rho_q^{(1)})]
     \sin^2(a_q^{(2)}\tau) \} \\
 n_{\downarrow}
 &=& \frac{1}{2\pi}
     \{\sin^2(\rho_q^{(1)})
    +[ \sin^2(2\rho_q^{(2)}
      -\rho_q^{(1)})
      -\sin^2(\rho_q^{(1)})]
     \sin^2(a_q^{(2)}\tau) \}.
\end{eqnarray}
The magnetization (or spin-polarization) is defined as
$P_z=(n_{\uparrow}-n_{\downarrow})/(n_{\uparrow}+n_{\downarrow})$, and we have
\begin{equation}
 P_z
  =  \cos(2\rho_q^{(1)})
    +[ \cos(4\rho_q^{(2)}
      -2\rho_q^{(1)})
      -\cos(2\rho_q^{(1)})]
     \sin^2(a_q^{(2)}\tau).
\end{equation}

This formula originates from a single-particle Hamiltonian. For N-particle
systems, when all the particles stay at the same state $\psi_{gs}$
initially and the interaction is neglected, it is straight forward to prove
that the above formula holds also. This formula gives a clear picture of a
harmonic $\theta$-indepensdent oscillation with a period $\tau_p=\pi/a_q^{(2)}$
which depends only on the second set of parameters, and with
an amplitude $A_{amp}=\cos(4\rho_q^{(2)}-2\rho_q^{(1)})-\cos(2\rho_q^{(1)})$.
Obviously, when $\rho_q^{(1)}=\pi/2$ and $\rho_q^{(2)}=\pi/4$,
$A_{amp}=2$ and the amplitude arrives at its maximum.
In general, when $\rho_q^{(1)}$ has been given and $\cos(2\rho_q^{(1)})\geq 0\
(<0)$, then $A_{amp}$ will arrive at its
conditional maximum at $\rho_q^{(2)}=\rho_q^{(1)}/2+\pi/4\ (\rho_q^{(1)}/2)$.
It implies that, when the first set of parameters are fixed,
one can tune the second set so that $A_{amp}$ is maximized. On the other
hand, when $\rho_q^{(2)}$ is close to $\rho_q^{(1)}$, $A_{amp}$ will
be very small. This is obvious from the expression of $A_{amp}$.

From the definition of the current, we obtain that the
up-current $j_{\uparrow}=j_{unit}n_{\uparrow}(q-\beta)$ and the
down-current $j_{\downarrow}=j_{unit}n_{\downarrow}(q+\beta)$,
where the unit of current is $j_{unit}=\hbar/(mR^2)$. The
currents are also oscillating with the same period
$\pi/a_q^{(2)}$, and they are also $\theta-independent$.
Obviously, when $|\beta|>|q|$, $j_{\uparrow}$ and
$j_{\downarrow}$ have different signs, and counter propagating
currents emerge.

\begin{figure}
 \resizebox{0.90\columnwidth}{!}{ \includegraphics{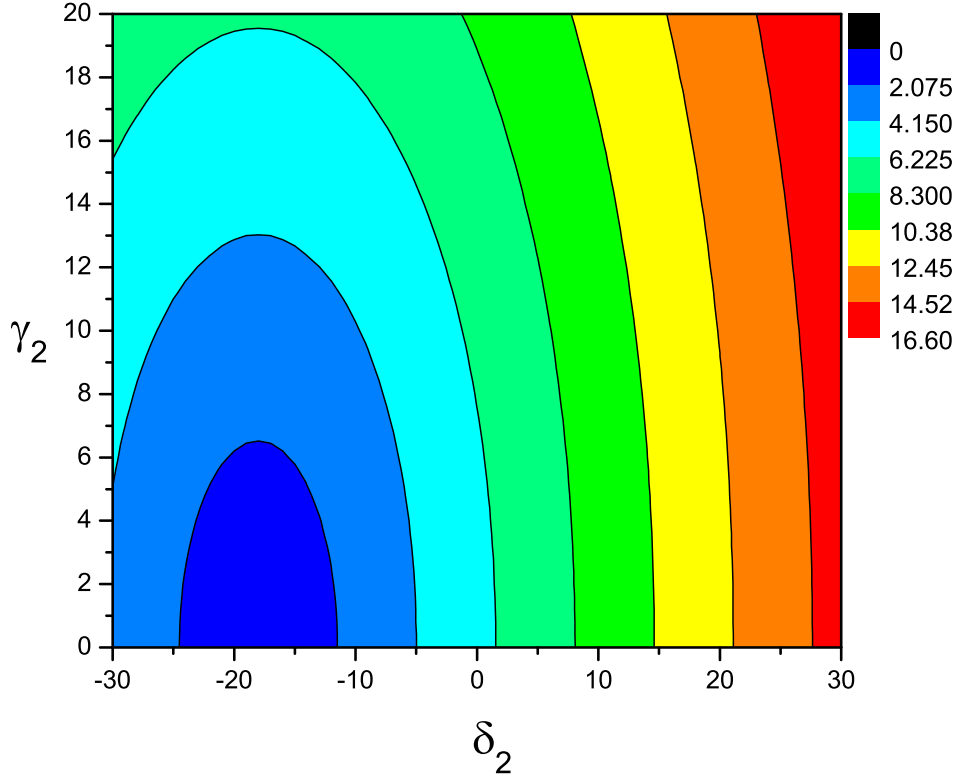} }
 \caption{The frequency $1/\tau_p$ versus $\delta_2$ and $\gamma_2$.
The oscillation is caused by a sudden change in $\gamma$ and $\delta$.
$\beta=3$ and $q=-3$ are given. The
minimum at which $1/\tau_p=0$ is located at $\delta_2=-18$ and $\gamma_2=0$.
The values of the contours are in a
arithmetic series. The contour closest to the right side has the largest
value $1/\tau_p=14.5$.}
\end{figure}

To give numerical results, the radius is given at $R=12\mu m$
in this paper. An example of the frequency
$1/\tau_p=a_q^{(2)}/\pi$ is shown in Fig.1. From the definition
of $a_q^{(2)}$ one can see that $1/\tau_{p}=0$ when
$\delta_2=2q\beta$ and $\gamma_2=0$. This minimum is shown at
the left-down corner of Fig.1 where $\delta_2=-18$. When
$\gamma_2$ increases from zero and/or $\delta_2$ goes away from
$2q\beta$, the frequency will increase.

\begin{figure}
 \resizebox{0.90\columnwidth}{!}{ \includegraphics{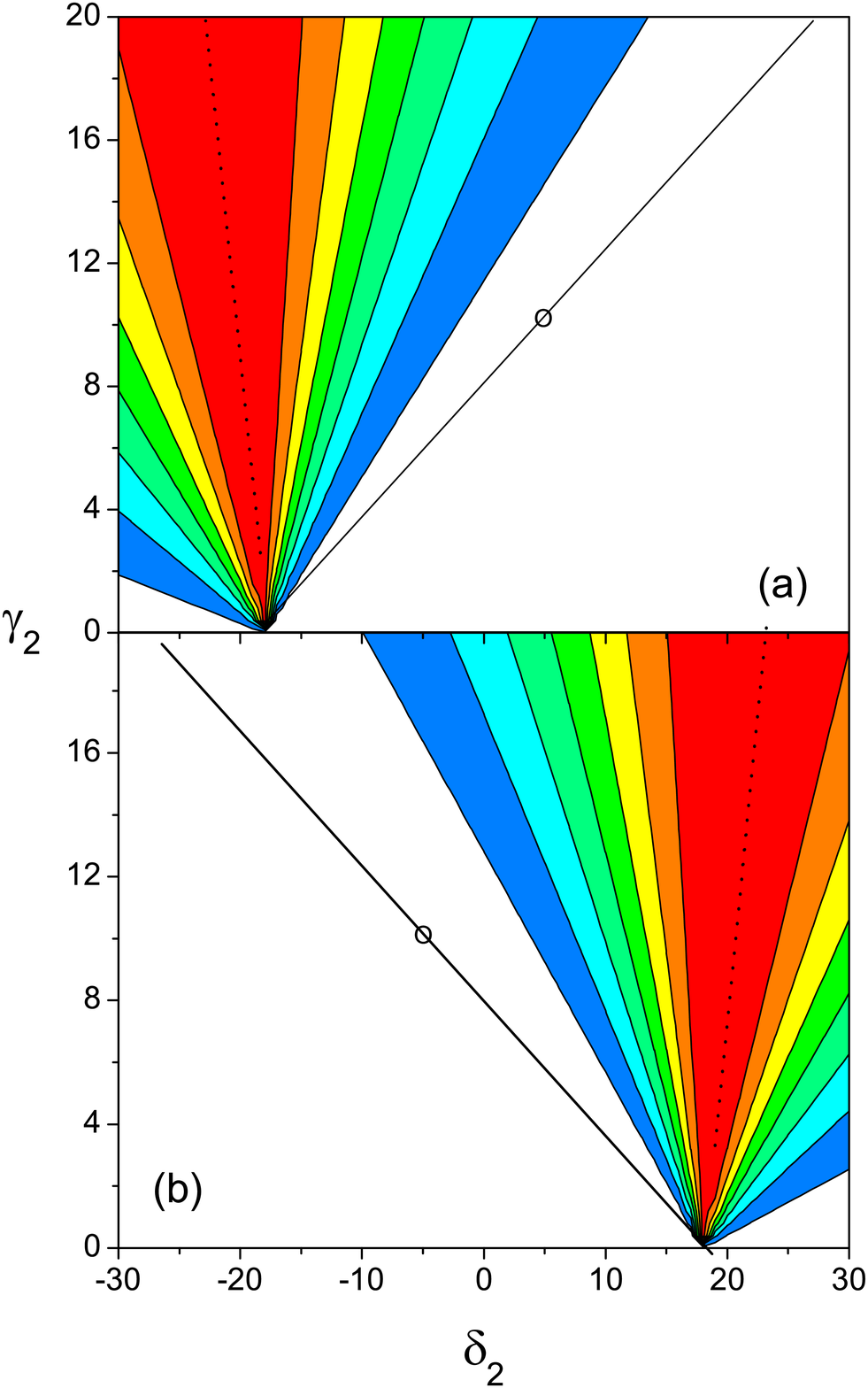} }
 \caption{The amplitude $A_{amp}$ versus $\delta_2$ and
$\gamma_2$. The oscillation is caused by a sudden change in
$\gamma$ and $\delta$. $\beta=3$, $\gamma_1={10}$, and
$\delta_1=5(-5)$, are given in a (b). Accordingly, the
ground state has $q=-3(3)$ in a (b). The dotted lines mark the conditional
maximum of $A_{amp}=1.917$ in both (a) and (b) (the largest value in each
panel). The contours going away from this line are in a arithmetic series
and decrease to zero. The black small circle marks the location where
$\delta_2=\delta_1$ and $\gamma_2=\gamma_1$. The solid line passing through
the circle marks the locations where $\rho_q^{(2)}=\rho_q^{(1)}$
and accordingly $A_{amp}=0$.}
\end{figure}

Two examples of $A_{amp}$ are shown in Fig.2. When $\gamma_2$ is small
and $\delta_2\simeq 2q\beta$, there is a narrow domain in which $A_{amp}$
is highly sensitive to $\delta _2$ and varies as a sharp peak versus
$\delta_2$. This is a distinguished feature. When $\gamma_2$ increases,
the width of the peak becomes broader. The summit of the peak is marked by a
dotted line.

Note that the first set of parameters are marked by a small circle in Fig.2.
Around it there is a broad domain in which $A_{amp}$\ is very small (i.e.,
in which $\rho_q^{(2)}\simeq\rho_q^{(1)}$). In particular, the
locations with $\rho_q^{(2)}=\rho_q^{(1)}$ are marked by a solid line
passing through the small circle. Along this line $A_{amp}=0$ and the sudden
change of the laser field can not cause an oscillation. Note that, when the
first set of parameters are fixed, $A_{amp}$ is determined by
$\rho_q^{(2)}$, and $\rho_q^{(2)}$ is in fact a function of
$|\gamma_2/(\delta_2-2q\beta)|$ together with the signs of
$\gamma_2$ and $\delta_2-2q\beta$. This explains that all the contours
in Fig.2 are straight lines.

Recently, in an experiment of cold atoms with the ring geometry,\cite{sb} by
making use of a radio frequency field, the initial state can be prepared in
a superposition state
\begin{equation}
 \psi_{init}
  =  \sin(\phi/2)
     \varphi_{q\uparrow}
    +\cos(\phi/2)
     \varphi_{q\downarrow},
\end{equation}
where $\phi$ is tunable ($0\leq\phi\leq\pi$) and determines the
initial ratio of the two components. $q$ is also tunable and is the initial
angular momentum of the particle. In this case we have the set of parameters
$\phi$, $q$, $\beta$, $\gamma$, and $\delta$. From Eq.(4) the
time-dependent solution is
\begin{eqnarray}
 \psi(\theta,t)
 &=& e^{-i\tau(q^2+2\beta^2)}
     ( f_{u_1}\varphi_{q,\uparrow}
      +f_{u_2}\varphi_{q-2\beta,\uparrow}
    +f_{d_1}
     \varphi_{q+2\beta,\downarrow}
    +f_{d_2}
     \varphi_{q,\downarrow}),
\end{eqnarray}
where
\begin{eqnarray}
 f_{u_1}
 &=& \sin(\phi/2)
     e^{-2iq\beta\tau}
     [ \cos(a_{q+\beta}\tau)
      +i
       \cos(2\rho_{q+\beta})
       \sin(a_{q+\beta}\tau) ], \\
 f_{u_2}
 &=&-i
     \cos(\phi/2)
     e^{2iq\beta\tau}
     \sin(2\rho_{q-\beta})
     \sin(a_{q-\beta}\tau), \\
 f_{d_1}
 &=&-i
     \sin(\phi/2)
     e^{-2iq\beta\tau}
     \sin(2\rho_{q+\beta})
     \sin(a_{q+\beta}\tau), \\
 f_{d_2}
 &=& \cos(\phi/2)
     e^{2iq\beta\tau}
     [ \cos(a_{q-\beta}\tau)
      -i
       \cos(2\rho_{q-\beta})
       \sin(a_{q-\beta}\tau) ].
\end{eqnarray}

\begin{figure}[tbp]
 \resizebox{0.90\columnwidth}{!}{ \includegraphics{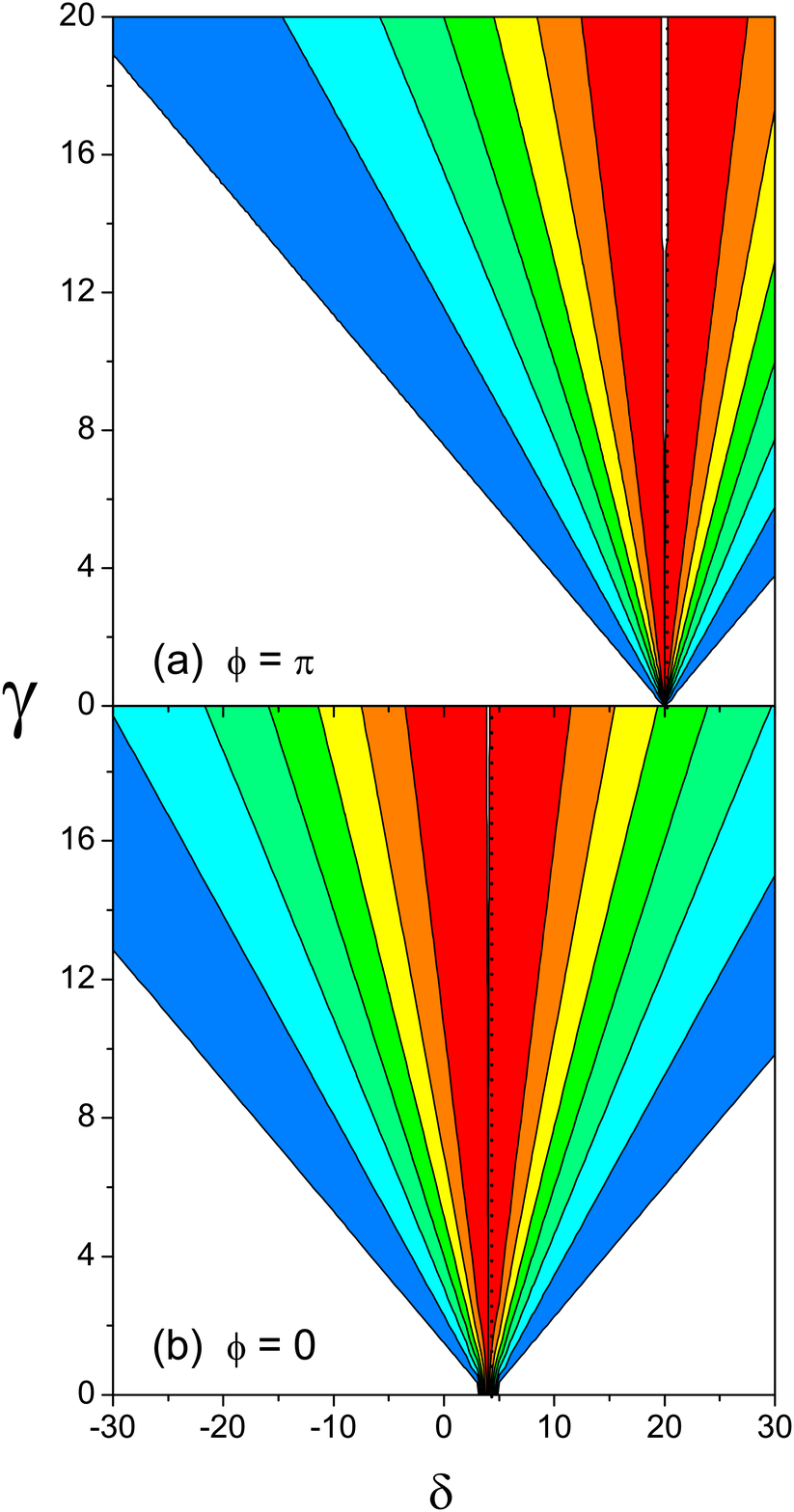} }
 \caption{The amplitude $A_{amp}$ versus $\delta$ and $\gamma $
for the case that the initial state is a superposition state.
$q=3$ and $\beta=2$ are assumed.  The dotted line marks the maximum of $A_{amp}=2 $. The
locations with $\gamma=0$ have $A_{amp}=0$. }
\end{figure}

Accordingly, we can obtain the up- and down-densities
$n_{\uparrow}(\theta,\tau)$ and $n_{\uparrow}(\theta,\tau)$ given
in the appendix. From them the magnetization $P_z$ can be obtained.
It turns out that, when $\phi=\pi $ or 0 (i.e., the atoms in
$\psi_{init}$ are purely up or purely down) the magnetization has
a very simple form as
\begin{eqnarray}
 P_z
 &=& \cos(4\rho_{q+\beta})
    +( 1
      -\cos(4\rho_{q+\beta}))
     \cos^2(a_{q+\beta}\tau),\ \ \ (\mbox{if }\phi=\pi), \\
 P_z
 &=&-\cos(4\rho_{q-\beta})
    -( 1
      -\cos(4\rho_{q-\beta}))
     \cos^2(a_{q-\beta}\tau),\ \ \ (\mbox{if }\phi=0).
\end{eqnarray}

As before, when all the particles stay in the same $\psi_{init}$ initially
and the interaction is neglected, the above formulae hold also for
N-particle systems. They give also a clear picture of harmonic oscillation,
but the periods are different for the two cases of $\phi $.

When $\phi=\pi$, $\tau_p=\pi/a_{q+\beta}$. If $\delta$ is tuned
so that $\delta=2\beta(q+\beta)\equiv\delta_o$, the frequency
$1/\tau_p$ will be minimized. When $\delta$ goes away from
$\delta_o$, the frequency increases. The amplitude
$A_{amp}=1-\cos(4\rho_{q+\beta})$. Obviously, when
$\rho_{q+\beta}=\pm\pi/4$, $A_{amp}$ arrives at its maximum 2.
This maximal oscillation can be achieved when
$\delta\rightarrow 2(q+\beta)\beta$ or
$|\gamma|\rightarrow\infty$. Whereas when $\gamma\rightarrow
0$, we always have $A_{amp}\rightarrow 0$ and the oscillation
does not appear. Thus $\gamma$ is the source of the
oscillation.

\begin{figure}[tbp]
 \resizebox{0.90\columnwidth}{!}{ \includegraphics{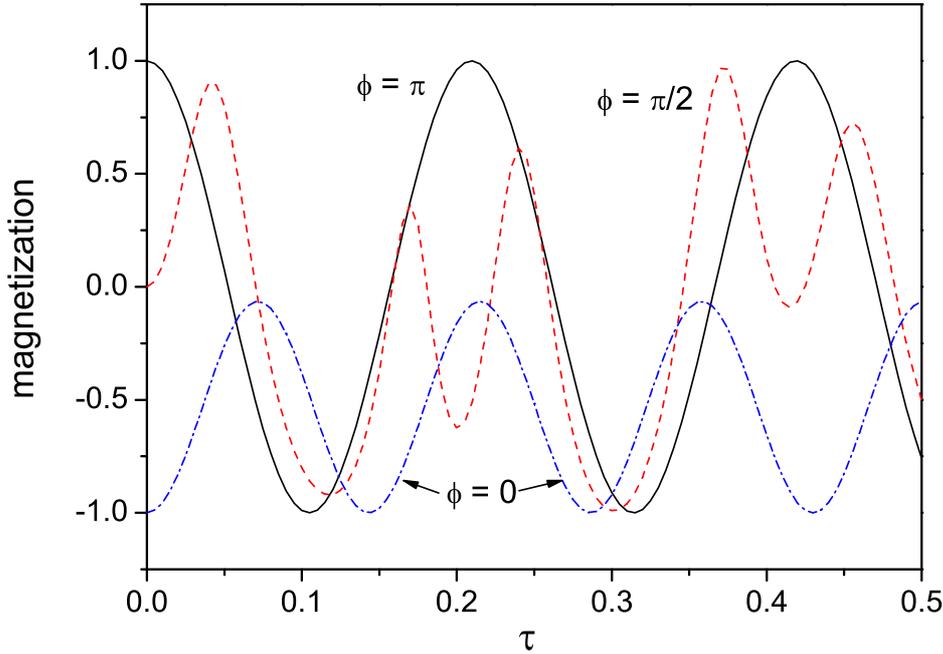} }
 \caption{The magnetization $P_z$ versus $\tau$ observed at
$\theta=0$. The initial state is a superposition
state with a $\phi$  marked by the curve. $q=3$, $\beta=2$,
$\gamma=15$, and $\delta=20$ are assumed. When $\phi=\pi$ or 0
the oscillation is harmonic. Otherwise, it is not. }
\end{figure}

When $\phi=0$, the discussion in the preceding paragraph holds also except
that $q+\beta$ should be changed to $q-\beta$. An example of $A_{amp}$ is
shown in Fig.3.

When $\phi$ is neither $\pi$ nor 0, the oscillation is no more
harmonic. In particular, both the up- and down-densities depend
on $\theta$ (refer to the appendix). Consequently, there are
stripes emerge along the ring, and the stripes move with time.
An example of the evolution of $P_z$ is shown in Fig.4. Where
the curve with $\phi=\pi$ has $A_{amp}=2$, while the curve with
$\phi=0$ has a much smaller $A_{amp}$ (the small amplitude is
caused by the choice of the parameters, refer to Fig.3). For
$\phi=\pi/2$, the anharmonic oscillation is shown.

\begin{figure}[tbp]
 \resizebox{0.90\columnwidth}{!}{ \includegraphics{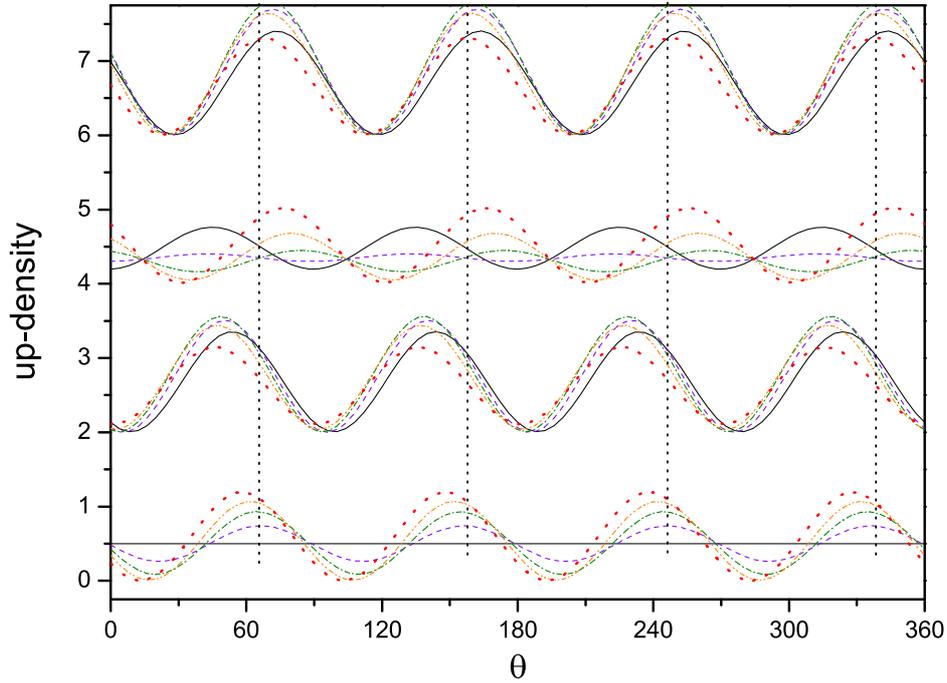} }
 \caption{Up-density $n_{\uparrow}$ versus $\theta$ with $\phi=\pi/2$.
The densities are given by 20 curves for a sequence of $\tau$. Starting
from $\tau=0$, in each step $\tau$ is increased by 1/40. The 20
curves in the sequence are divided into 4 groups. The $n_{\uparrow}$ of
each group has been shifted up by 2 relative to its preceding group, this is
just for guiding the eyes. The five $n_{\uparrow}$ in each group are
plotted in solid, dash, dash-dot, dash-dot-dot, and dot lines, respectively,
according to the time-sequence (say, the one with $\tau=0$ is the horizontal
solid line in the lowest group). The parameters are $q=1$, $\beta=2$,
$\gamma=10$, and $\delta=1$. The vertical dotted lines are also for guiding the eyes.}
\end{figure}

An example of the up-density with $\phi=\pi/2$ is shown in Fig.5, where
the stripes emerging along the ring are shown. Note that two waves are
contained in each component (say, the up-component is a mixture of
$\varphi_{q,\uparrow}$ and $\varphi_{q-2\beta,\uparrow}$). The stripes arise
from the interference of these two waves. The number of peaks (valleys) is
$2\beta$ (say, the number is four in Fig.5). Their locations and clarity
vary with time.

When the interaction is taken into account, the total Hamiltonian is
$H=\sum_i\hat{h}_i+\sum_{i<j}V_{ij}$, where
$V_{ij}=g\delta(\theta_i-\theta_j)$, $g=g_{\upuparrows}$ (if both atoms are
up), $g_{\downdownarrows}$ (both down), or $g_{\uparrow\downarrow}$ (one
up and one down). In order to evaluate the effect of interaction in a
simplest way, we study a two-body system. Firstly, the set of single
particle states $\varphi_{k\uparrow}$ and $\varphi_{k\downarrow}$ are
adopted. A constraint $-k_{\max}\leq k\leq k_{\max}$ is set. Accordingly,
we have $2(2k_{\max}+1)$ single particle states, and they are renamed as
$\varphi_j\equiv\varphi_{k_j\mu_j}$ ($\mu_j=\uparrow$ or $\downarrow$).
Based on $\varphi_j$ a set of basis functions for the
two-body system $\Phi_i=\tilde{S}[\varphi_j(1)\varphi_{j'}(2)]$
are defined, where $\tilde{S}$ is for
the symmetrization and normalization, and $j\geq j'$.

\begin{figure}[htbp]
 \resizebox{0.90\columnwidth}{!}{ \includegraphics{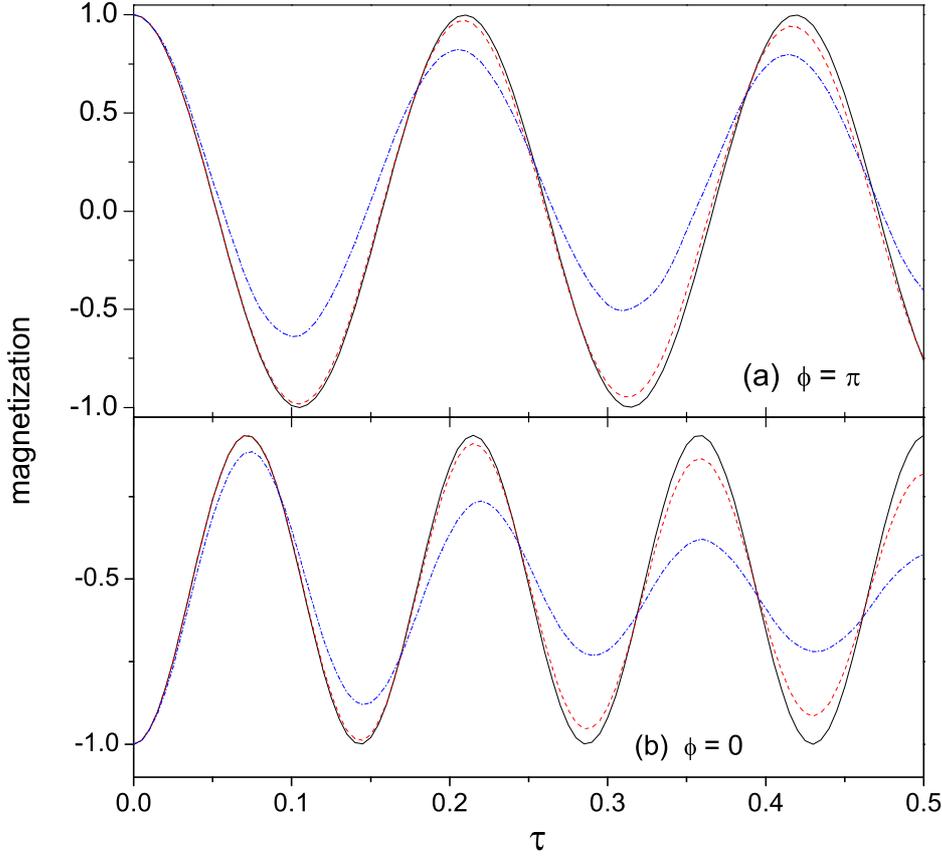} }
 \caption{$P_z$ of a 2-body system of Rb atoms versus $\tau$ with
interaction. $r_w=R/\sqrt{18}$ and
$g_{\uparrow\uparrow}=g_{\downarrow\downarrow}=g_{\uparrow\downarrow}=g$ are
assumed. The solid, dash, and dash-dot curves have $g=0$, $100g_{eff}$, and
$2000g_{eff}$, respectively. The initial state has $\phi=\pi$ (a) and
$=0$ (b). The other parameters are the same as in Fig.4 }
\end{figure}

\

\

\

Note that the strength of a pair of realistic Rb atoms is $g_{Rb}=7.79\times
10^{-12}H_Z cm^3$ (the differences in the strengths between the up-up,
up-down, and down-down pairs are very small and are therefore neglected).
Since the atoms in related experiments are not really distributed exactly on
a one-dimensional ring but in a domain surrounding the ring, the effect of
the diffused distribution should be considered. Hence, for each
$\frac{1}{\sqrt{2\pi}}e^{ik\theta}$, we define its 3-dimensional counterpart
$\frac{1}{\pi r_w}\sqrt{\frac{1}{R}}e^{ik\theta}e^{-(r/r_w)^2}$, where the
Gaussian function $e^{-(r/r_w)^2}$ describes the diffused distribution
and $r_w$ measures the width of the distribution. Accordingly, for each
$\varphi_{k\mu}$, we have its counterpart $\varphi_{k\mu}^{[3d]}$. With
$\varphi_{k\mu}^{[3d]}$ we define an effective strength $g_{eff}$ so that
for any pair of matrix elements

\begin{eqnarray}
 & & g_{eff}
     \int d\theta_i d\theta_j
     \varphi_{k_1\mu_1}^{\dagger}
     \varphi_{k_2\mu_2}^{\dagger}
     \delta(\theta_i-\theta_j)
     \varphi_{k_3\mu_3}
     \varphi_{k_4\mu_4}  \nonumber \\
 &=& g_{Rb}
     \int d\mathbf{r}_i d\mathbf{r}_j
     \varphi_{k_1\mu_1}^{[3d]\dagger}
     \varphi_{k_2\mu_2}^{[3d]\dagger}
     \delta(\mathbf{r}_i-\mathbf{r}_j)
     \varphi_{k_3\mu_3}^{[3d]}
     \varphi_{k_4\mu_4}^{[3d]}.
\end{eqnarray}

Then, we have $g_{eff}=\frac{1}{\pi R\ r_w^2}g_{Rb}$ which is the
strength adopted in our calculation.

It is assumed that each of the two atoms is given in a superposition state
with either $\phi=\pi$ or 0 initially. When the Hamiltonian is
diagonalized in the space expanded by $\Phi_i$, the eigenenergies $E_l$
and eigenstates $\Psi_l\equiv\sum_i C_{li}\Phi_i$ can be
obtained, and the time-dependent state is
\begin{equation}
 \Psi(\theta,t)
  =  \sum_l
     e^{-i\tau E_l}
     \Psi_l\langle
     \Psi_l |
     \Psi_{init}\rangle.
\end{equation}
From $\Psi(\theta,t)$ the densities and $P_z(t)$ can be
calculated. An examples with $k_{\max}=14$ is shown in Fig.6.
When $k_{\max}$ is changed from 14 to 12, there is no explicit
changes in the pattern. It implies that the choice
$k_{\max}=14$ is sufficient in qualitative sense. In Fig.6 the
curves "1" and "2" overlap nearly. It implies that the effect
of interaction with a $g<100g_{eff}$ is very small. Thus the
effect of interaction for the two-body system with $R=12\mu m$
is negligible. However, when $g=2000g_{eff}$, the amplitude
decreases with time explicitly as shown by the dash-dot curve,
while the period remains nearly unchanged.

Note that the effect of interaction depends also on the particle density.
For a  N-body system with a larger $N$\ (or the ring becomes smaller) the
effect would become stronger. The phenomenon shown by the dash-dot curve has
already been observed in existing experiments for N-body condensates.\cite{jyz,yjl1}
It is possible that the qualitative features of the oscillations
of a many-body system and a few-body system with a stronger strength would
be more or less similar. This is a topic to be studied further.

In summary, the oscillation of the cold atoms under the qSO and constrained
on a ring is studied analytically for arbitrary $N$ without interactions.
Then, the effect of the interaction is evaluated numerically via a two-body
system. Two cases, namely, the evolution starting from a ground state
induced by a sudden change of the laser field, and the evolution starting
from a superposition state, have been studied. The emphasis is placed on
clarifying the relation between the parameters of the laser beams (causing
the qSO) and the period and amplitude of the oscillation. This is achieved
by giving a set of formulae so that the relation can be understood
analytically. It has been predicted that, under certain conditions, the
oscillation can be maximized or minimized, and oscillating
counter-propagating currents will emerge. Experimental confirmation of the
regularity unveiled in this paper is expected.

\acknowledgments
The support from the NSFC (China) under the grant number 10874249 is
appreciated.

\

\

\

\textbf{APPENDIX}

When the initial state is $\psi_{init}=\sin(\phi/2)\varphi_{q\uparrow}
+\cos(\phi/2)\varphi_{q\downarrow}$, the associated up- and
down-densitues during the evolution are
\begin{eqnarray}
 n_{\uparrow}(\theta,\tau)
 &=& \frac{1}{2\pi}
     \{ \sin^2(\phi/2)
        [ \cos^2(a_{q+\beta}\tau)
         +\cos^2(2\rho_{q+\beta})
          \sin^2(a_{q+\beta}\tau) ] \nonumber \\
  &&+\cos^2(\phi/2)
     \sin^2(2\rho_{q-\beta})
     \sin^2(a_{q-\beta}\tau) \nonumber \\
  &&-\sin\phi\sin(2\rho_{q-\beta})
     \sin(a_{q-\beta}\tau)
     [ \cos(2\beta\theta-4q\beta\tau)
       \cos(2\rho_{q+\beta})
       \sin(a_{q+\beta}\tau) \nonumber \\
  &&+\sin(2\beta\theta-4q\beta\tau)
     \cos(a_{q+\beta}\tau) ] \}, \\
 n_{\downarrow}(\theta,\tau)
 &=& \frac{1}{2\pi}
     \{ \cos^2(\phi/2)
        [ \cos^2(a_{q-\beta}\tau)
         +\cos^2(2\rho_{q-\beta})
          \sin^2(a_{q-\beta}\tau ) ] \nonumber \\
  &&+\sin^2(\phi/2)
     \sin^2(2\rho_{q+\beta})
     \sin^2(a_{q+\beta}\tau) \nonumber \\
  &&+\sin\phi
     \sin(2\rho_{q+\beta})
     \sin(a_{q+\beta}\tau)
     [ \cos(2\beta\theta-4q\beta\tau)
       \cos(2\rho_{q-\beta})
       \sin(a_{q-\beta}\tau) \nonumber \\
  &&+\sin(2\beta\theta-4q\beta\tau)
     \cos(a_{q-\beta}\tau) ] \}.
\end{eqnarray}

\end{document}